\documentclass[preprint,superscriptaddress]{paper}
\usepackage{a4wide}
\usepackage{textcomp}
\usepackage{amssymb}
\usepackage{graphics}
\usepackage{graphicx}
\usepackage{epsfig}
\usepackage{amsthm}

\begin{document}

\title{A self-consistent approach to the one-dimensional chain models for
the ferro- and antiferro-magnetism of nanotubes}

\author{Zhaosen Liu$^{a}$, Hou Ian$^b$}
\institution{$^a$Department of Applied Physics, Nanjing University of Information \\ Science and Technology, Nanjing 210044, China \\
$^b$ Institute of Applied Physics and Materials Engineering, FST, \\ University of Macau, Macau}

\maketitle
\begin{abstract}
We employ a self-consistent simulation approach based on quantum theory
to investigate the physical properties of a pair of ferromagnetic
and antiferromagnetic nanotubes. It was observed that under the given
conditions, no matter the external magnetic field was absent or applied
along the easy longitudinal axis, the spins always ordered in that
direction due to the special geometric shape of the tubes and the
magnetic uniaxial anisotropy, so that the two sorts of nanosystems
exhibit typical ferromagnetic and antiferromagnetic properties, which
may find applications in modern technology, in strong contrast to
the phenomena observed previously in nanoparticle \cite{liuphye12},
where an external magnetic field, applied parallel to the antiferromganetically
coupled spins, is able to turn the spins off their original direction
to form symmetric pattern around the field direction. This peculiar
feature enable us to build up one-dimensional chain models for the
two sorts of nanosystems. Considering their fast computational speed
and simplicity, these theoretical models were then utilized to investigate
the finite size effects of the nanosystems, and perform further analysis
for long tubes. Especially, our results obtained with the theoretical
models and the numerical approach are exactly identical, verifying
the correctness and applicability of the computational methodology. 
\end{abstract}
%\pacs{75.40.Mg,  Numerical simulation studies%75.10.Jm  % Quantized spin models, including quantum spin frustration}

Since the discovery of carbon nanotubes \cite{Iijima}, the field
of this kind of nanosystems, especially those exhibiting magnetic
properties, have been attracting considerable attention due to their
potential applications in modern technology as well as their significance
in science \cite{HLiu1,HLiu2,Fagan,HLiu3,HLiu4}. For instance, high
attention has been paid to the magnetic nanotubes based on the transition
metals, such as CoPt, CoPd, FePt and FePd alloys owing to their potential
applications in high-density magnetic recording media \cite{Gao2,Peng,Hu,Schaaf,Takata}.
Now, one can find various new nanotubes based electronic and spintronic
devices \cite{Krusin-Elbaum}, multi-functional systems involving
nanotubes for biotechnological applications \cite{Gao}.

So far, many techniques for fabricating magnetic nanotubes have been
developed \cite{Jang,Xie,Kim,Ding,Bogani,Gul}. Hematite $\alpha$-Fe$_{2}$O$_{3}$
nanotubes can be produced for example by using a template method,
which provides nanotubes with diameters between 100 nm and 200 nm,
and a wall thickness less than 10 nm \cite{Xie}; attachment of maghemite
$\gamma$-Fe$_{2}$O$_{3}$ nanoparticles to the surface of carbon
nanotubes via a modified sol-gel technique has also produced nano
hybrid structures behaving as nanotubes \cite{Kim}.

Theoretically, electronic structures of ferromagnetic single-wall
nanotubes were investigated by means of spin density functional theory
\cite{Shimada}; continuous models for ferromagnetism \cite{Landeros},
approaches based on Green functions \cite{Mi}, and simulation methods
like Monte Carlo \cite{Masrour}, have been also employed to investigate
the exotic magnetic properties and spin configurations of magnetic
nanotubes.

In the present work, we first investigate the magnetic properties
of a pair of ferromagnetic and antiferromagnetic nanocylinders, in
which Heisenberg and uniaxial anisotropy along the central longitudinal
$z$-axis co-exist, by means of a self-consistent approach based on
quantum mean field theory. We find that the special geometric shape
of the tube surfaces and the uniaxial anisotropy force the spins order
in the easy $z$ direction, so that the two sorts of nanosystems exhibit
typical ferromagnetic and antiferromagnetic properties, which may
find applications in modern technology, in strong contrast to the
phenomena observed in antiferromagntic nanoparticle \cite{liuphye12},
where an external magnetic field, parallel to the oppositely aligning
spins, is able to turn the spins off their original direction to form
symmetric pattern around the field direction. This peculiar feature
of the nanosystems enables us to build up one-dimensional ferromagnetic
and antiferromagnetic chain models for theoretical analysis. It turns
out that the magnetizations, hysteresis curves, etc., generated by
means of the numerical approach and the theoretical models are exactly
identical, verifying our computational methodology once again.

\section{Computational Model}

Now, we consider a nano-cylinder, which is formed by rolling up a
rectangular monolayer lattice consisting of $S$ =1 spins sitting
on the square lattice sites. The length of the nanotube is ${\cal L}=N_{L}a$
along the longitudinal central axis, where $a$ is the lattice parameter,
and there are $N_{R}$ spins on every circle around the axis, i.e.,
the circumference of such a circle is ${\cal C}=N_{R}a$. Thus the
coordinators of a spin on the nanotube wall are determined by 
\begin{eqnarray}
x_{n} & = & \frac{N_{R}a}{2\pi}\cos\left(\frac{2\pi n}{N_{R}}\right)\nonumber \\
y_{n} & = & \frac{N_{R}a}{2\pi}\sin\left(\frac{2\pi n}{N_{R}}\right),\\
z_{m} & = & ma\;,\nonumber 
\end{eqnarray}
where $1\leq n\leq N_{R}$, and $1\leq m\leq N_{L}$.

To do Monte Carlo simulations for Ising-like nanocylinders, Masrour
et.~al. only considered Heisenberg exchange interaction among the
nearest neighboring spins with uniform strength, the uniaxial anisotropy
along the longitudinal central axis, and the external magnetic field
exerted in that direction \cite{Masrour}. Following Masrour's work,
the Hamiltonian for our single wall nanotubes thus can be expressed
as 
\begin{equation}
{\cal H}=-\frac{1}{2}\sum_{i,j\neq i}{\cal J}_{ij}{\vec{S}_{i}\cdot}\vec{S}_{j}-K_{A}\sum_{i}S_{iz}^{2}-g_{S}\mu_{B}\sum_{i}S_{iz}B\;,\label{hamlt}
\end{equation}
where $g_{s}$=2 is the Land{é} factor, ${\cal J}_{ij}$ and $K_{A}$
represent the strengths of the exchange interaction among the neighboring
spins and the uniaxial anisotropy assumed to be in the $z$-direction
as well. In the above Hamiltonian, the spins are quantum operators
instead of the classical vectors. Since $S=1$, the matrices of the
three spin components in Cartesian coordinate system are given by
{\footnotesize{}
\begin{eqnarray}
S_{x}=\frac{1}{2}\left(\begin{array}{ccc}
0 & \sqrt{2} & 0\\
\sqrt{2} & 0 & \sqrt{2}\\
0 & \sqrt{2} & 0
\end{array}\right)\;, & S_{y}=\frac{1}{2i}\left(\begin{array}{ccc}
0 & \sqrt{2} & 0\\
-\sqrt{2} & 0 & -\sqrt{2}\\
0 & \sqrt{2} & 0
\end{array}\right)\;, & S_{z}=\left(\begin{array}{ccc}
1 & 0 & 0\\
0 & 0 & 0\\
0 & 0 & -1
\end{array}\right)\;,
\end{eqnarray}
} respectively.

Since only the exchange interactions between the nearest spins are
considered, a spin on the tube wall interacts only with its two nearest
neighbors in the same line parallel to $z$ axis with a strength of
${\cal J}_{1}$, two nearest neighbors on the same circle around the
central axis with a strength of ${\cal J}_{2}$. When both ${\cal J}_{1}$
and ${\cal J}_{2}$ are positive, the nanotube is ferromagnetic, the
spins tend to align in the same direction; while both ${\cal J}_{1}$
and ${\cal J}_{2}$ are negative, the nanotube is antiferromagnetic,
every pair of neighboring spins attempt to order in the opposite directions.

This Hamiltonian given by Eq.(\ref{hamlt}) was first used to study
the magnetic properties of the nanotubes numerically by means of the
SCA approach \cite{liuphye12,liujpcm,liupssb}. In every simulation
step, it was diagonalized so as to calculate the magnetic moment of
the considered spin. All of our simulations were started here from
a random magnetic configuration and a temperature well above the magnetic
transition, then carried out stepwise down to very low temperatures
with a temperature step $\Delta T<0$. At any temperature, if the
difference $|\langle\vec{S}'_{i}\rangle-\langle\vec{S}_{i}\rangle|$
between two successive iterations for every spin is less than a very
small given value $\tau_{0}$, convergency is considered to be reached
\cite{liuphye12,liujpcm,liupssb}.

\section{Results Obtained Numerically with the SCA Approach}

%%%%%%%%%Fig1%%%%%%%%%%%%%%%
\begin{figure*}[h]
\centerline{\epsfig{file=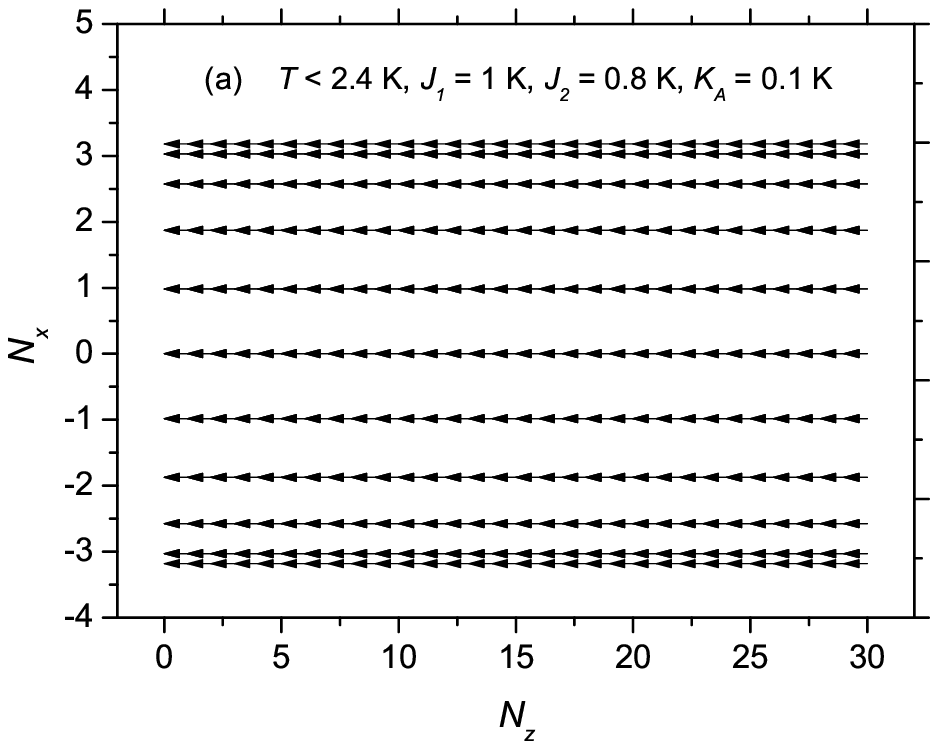,width=.55\textwidth,height=5cm,clip=}
\epsfig{file=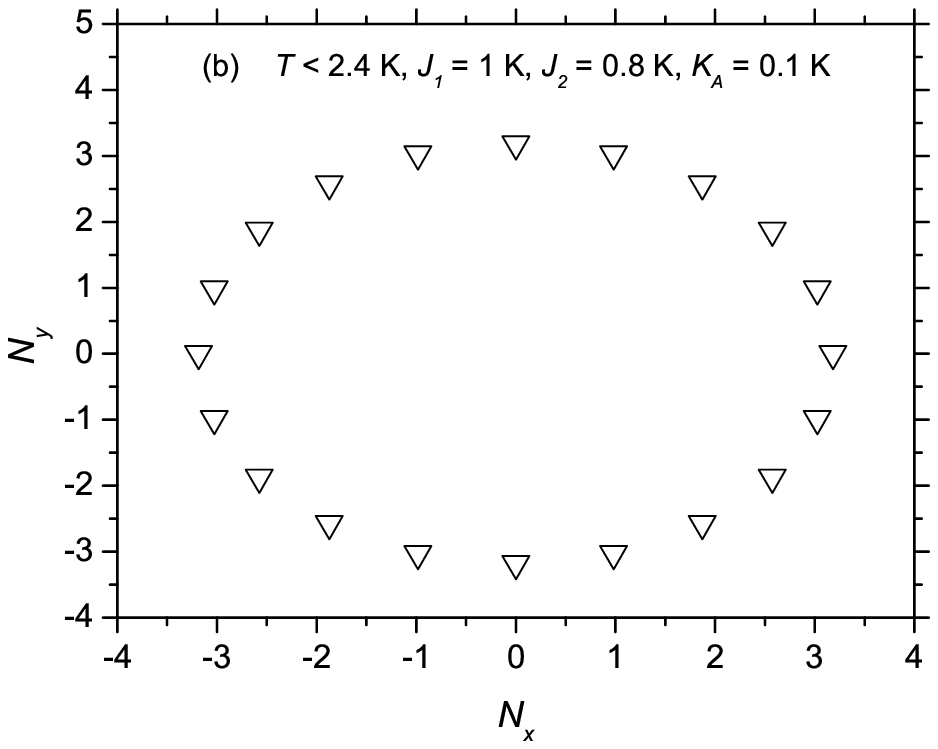,width=5cm,height=5cm,clip=}} \centerline{\epsfig{file=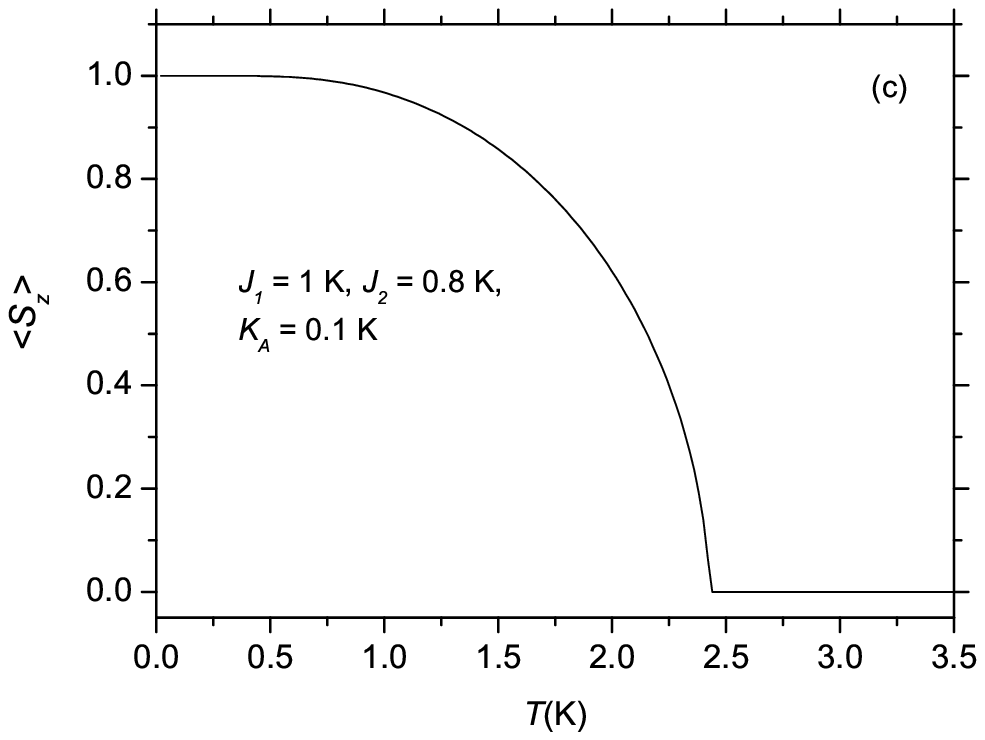,width=.45\textwidth,height=6.5cm,clip=}
\epsfig{file=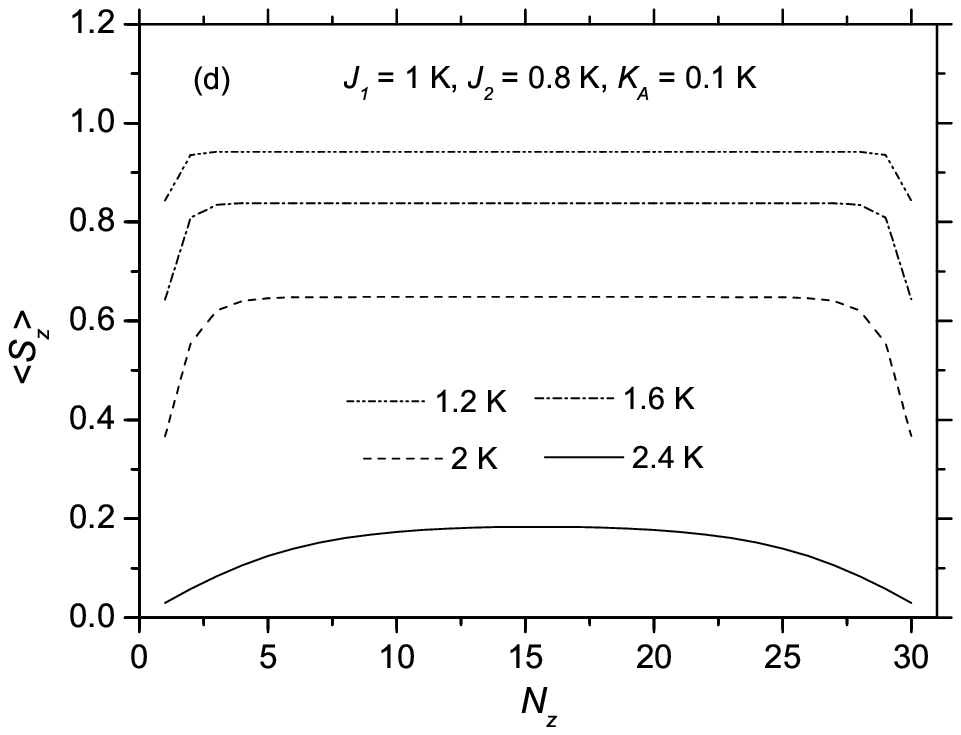,width=.45\textwidth,height=6.5cm,clip=}} 

\centering{}%
\parbox[c]{16cm}{%
{\small{}{}\textbf{\small{}Figure 1.}{\small{} Simulated magnetic
structures projected onto (a) $y$ = 0, and (b) $z$ = 15$a$ cross
sections, respectively; calculated $\langle S_{z}\rangle$ (c) as
function of temperature, and (d) as the function of $z$ at different
temperatures, for the ferromagnetic nanotube by means of the SCA approach
in the absence of external magnetic field. Here $N_{L}$ = 30. }}%
} 
\end{figure*}

Firstly, we performed simulations for a ferromagnetic nanotube with
$N_{L}$ = 30 and $N_{R}$ = 20 by means of the SCA approach. To do
this, Heisenberg exchange and uniaxial anisotropy strengths were assigned
to ${\cal J}_{1}$ = 1 K, ${\cal J}_{2}$ = 0.8 K and $K_{A}$ = 0.1
K, respectively. Fig.1(a,b) displays the magnetic structures projected
onto the $y$ = 0 and $z$ = 15$a$ cross sections. As shown there,
all spin order ferromagnetically antiparallel to the $z$-axis below
Curie temperature $T_{C}\approx$ 2.44 K. That is, in the magnetic
phase, only $\langle S_{z}\rangle$ is nonzero, both $\langle S_{x}\rangle$
and $\langle S_{y}\rangle$ all vanish. As displayed in Fig.1(c),
while temperature arises, $\langle S_{z}\rangle$ decreases gradually
from the maximum value at very low temperatures to zero until $T_{C}$.
Due to the finite length of the nanotube, we can find from Fig.1(d)
that the spins near the two ends of the nanotube naturally have weaker
magnitudes than those in the middle part at all recorded temperatures.

%%%%%%%%%Fig2%%%%%%%%%%%%%%%
\begin{figure*}[h]
\centerline{\epsfig{file=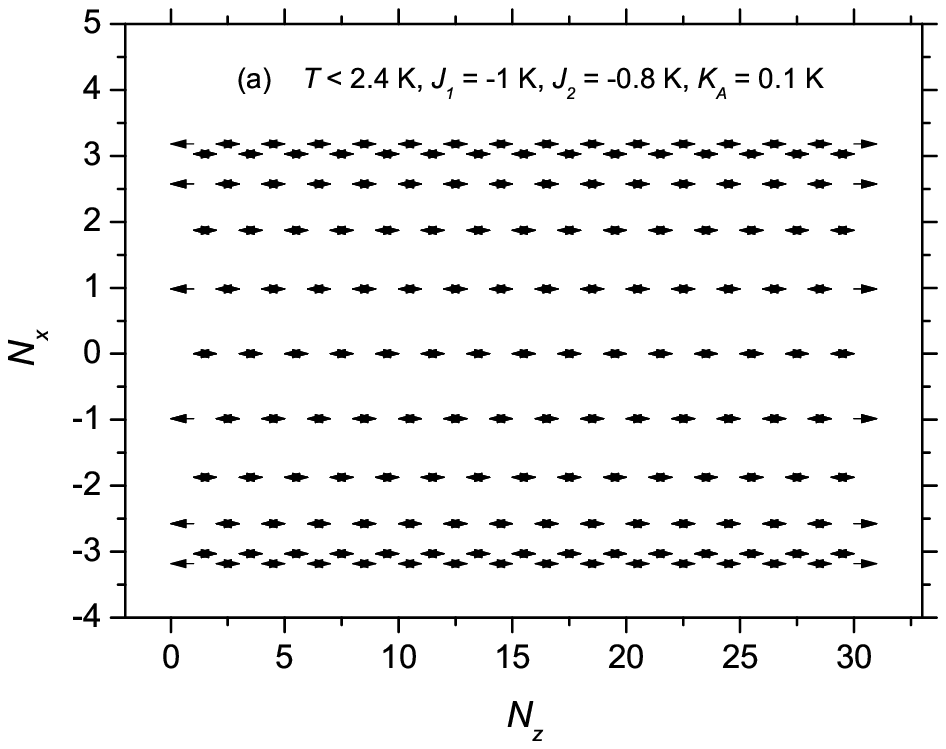,width=.55\textwidth,height=5cm,clip=}
\epsfig{file=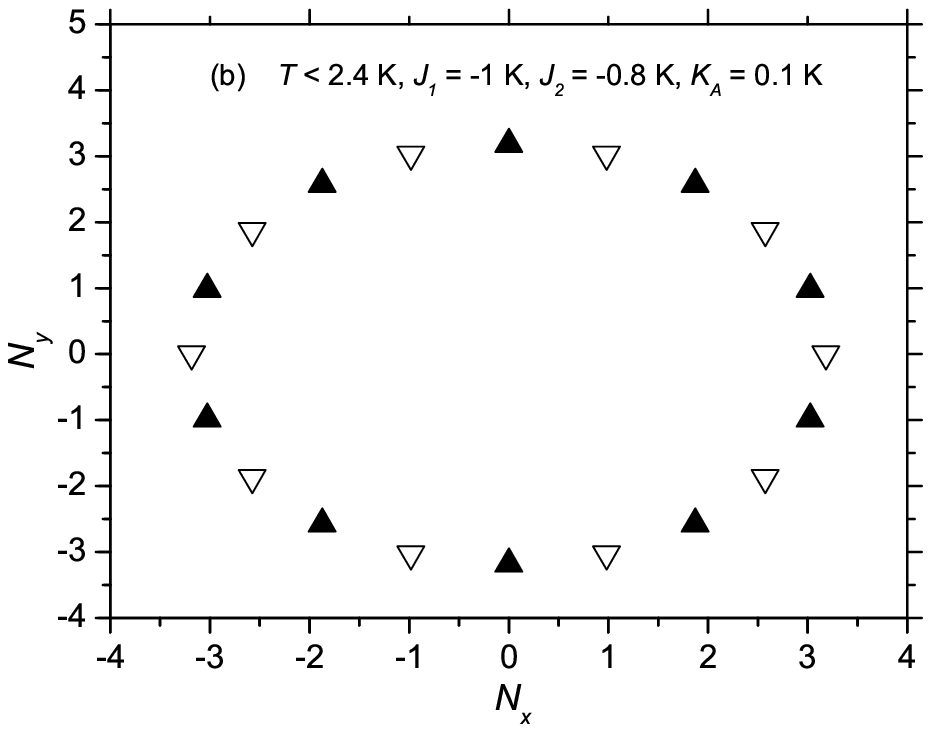,width=5cm,height=5cm,clip=}} \centerline{\epsfig{file=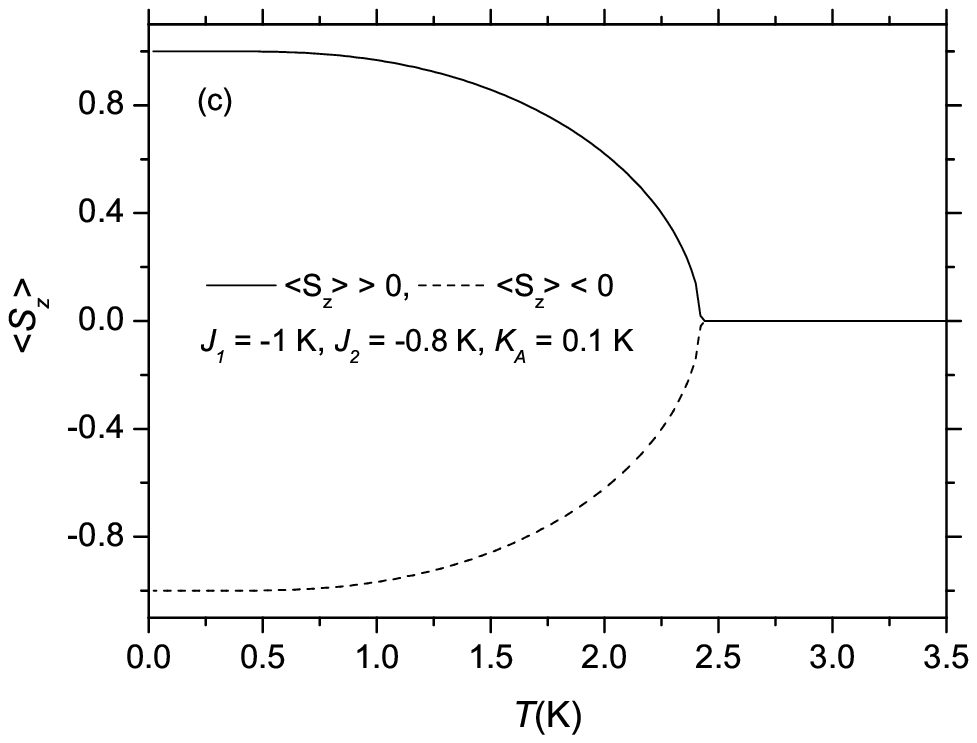,width=.45\textwidth,height=6.5cm,clip=}
\epsfig{file=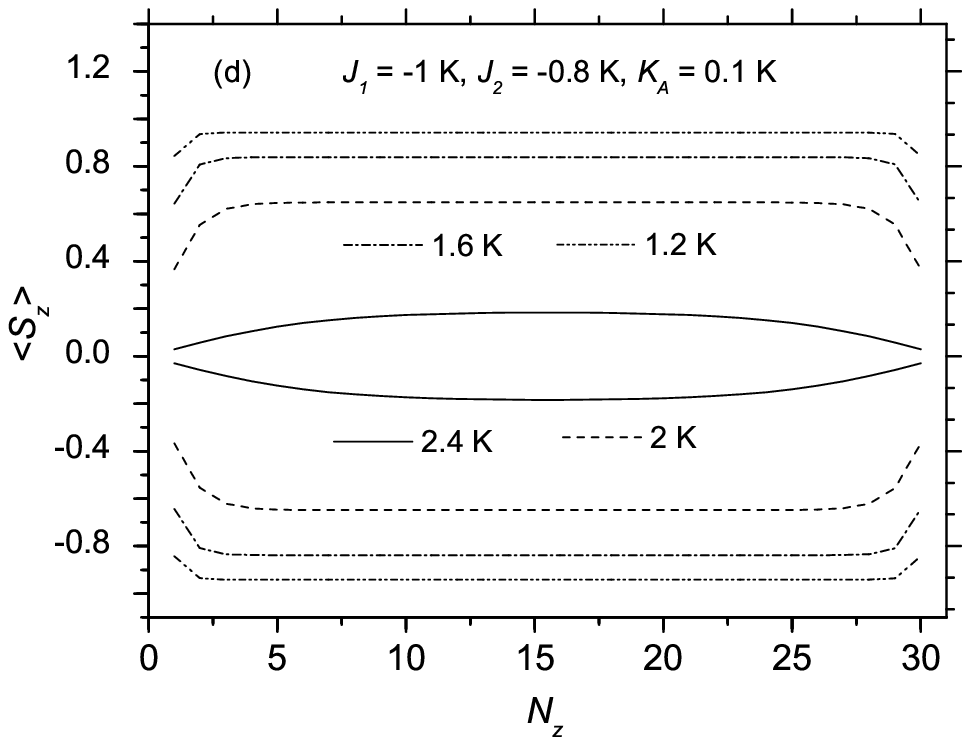,width=.45\textwidth,height=6.5cm,clip=}} 

\centering{}%
\parbox[c]{16cm}{%
{\small{}{}\textbf{\small{}Figure 2.}{\small{} Simulated magnetic
structures projected onto (a) $y$ = 0, and (b) $z$ = 15$a$ cross
sections; calculated $\langle S_{z}\rangle$ (c) as the function of
temperature, and (d) as the function of $z$ at different temperatures,
for the antiferromagnetic nanotube by means of the SCA approach in
the absence of external magnetic field. Here $N_{L}$ = 30. }}%
} 
\end{figure*}

For comparison with the results just obtained, we carried out simulations
for an antiferromagnetic nanotube of the same size scale with the
SCA approach by only changing the signs of the Heisenberg exchange
constants, that is, now ${\cal J}_{1}$ = -1 K and ${\cal J}_{1}$
= -0.8 K. Fig.2(a,b) depict the magnetic configurations projected
onto the $y$ = 0 and $z$ = 15$a$ cross sections. There we find
that all spin order antiferromagnetically in the $z$-direction below
Néel temperature $T_{N}\approx$ 2.44 K. Once again, in the magnetic
phase, only $\langle S_{z}\rangle$ is nonzero, both $\langle S_{x}\rangle$
and $\langle S_{y}\rangle$ all vanish. As temperature arises, the
two opposite $\langle S_{z}\rangle$'s, depicted in Fig.2(c), attenuate
gradually from the saturated value $S$ =1 at very low temperatures
to zero at $T_{N}$. As $N_{z}$ approaches to 1 or $N_{L}$ from
the middle region, $|\langle S_{z}\rangle|$ decreases as seen in
Fig.2(d). However, these curves exhibit excellent symmetry both horizontally
and vertically.

%%%%%%%%%Fig3%%%%%%%%%%%%%%%
\begin{figure*}[ht]
\centerline{\epsfig{file=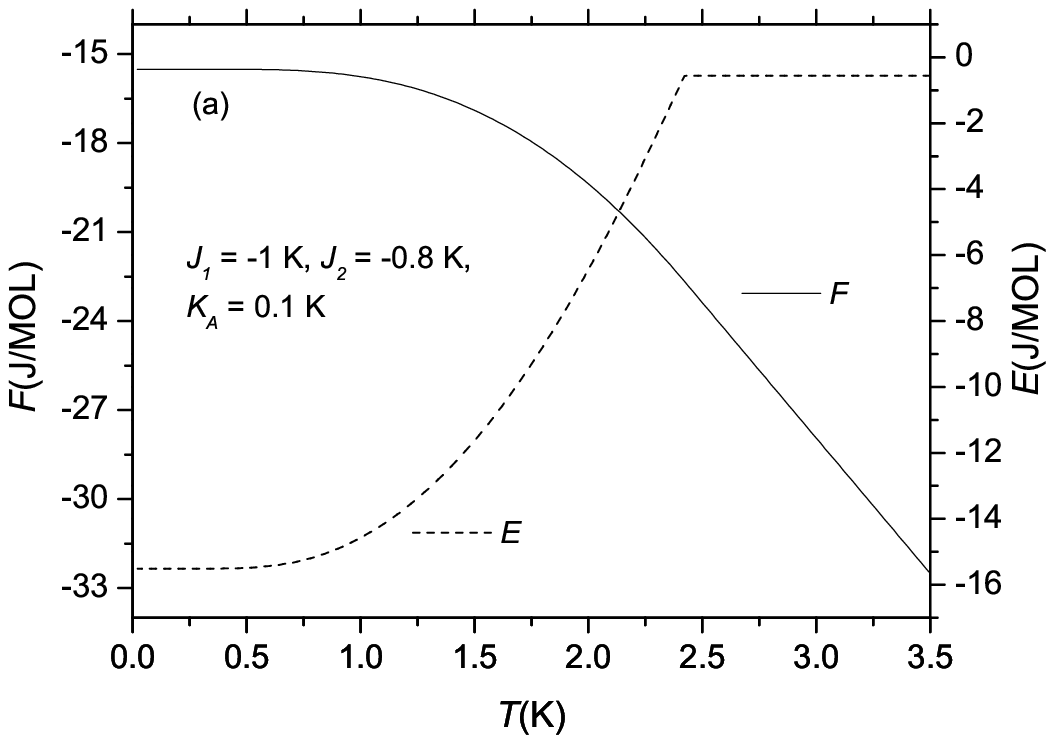,width=.45\textwidth,height=6.5cm,clip=}
\epsfig{file=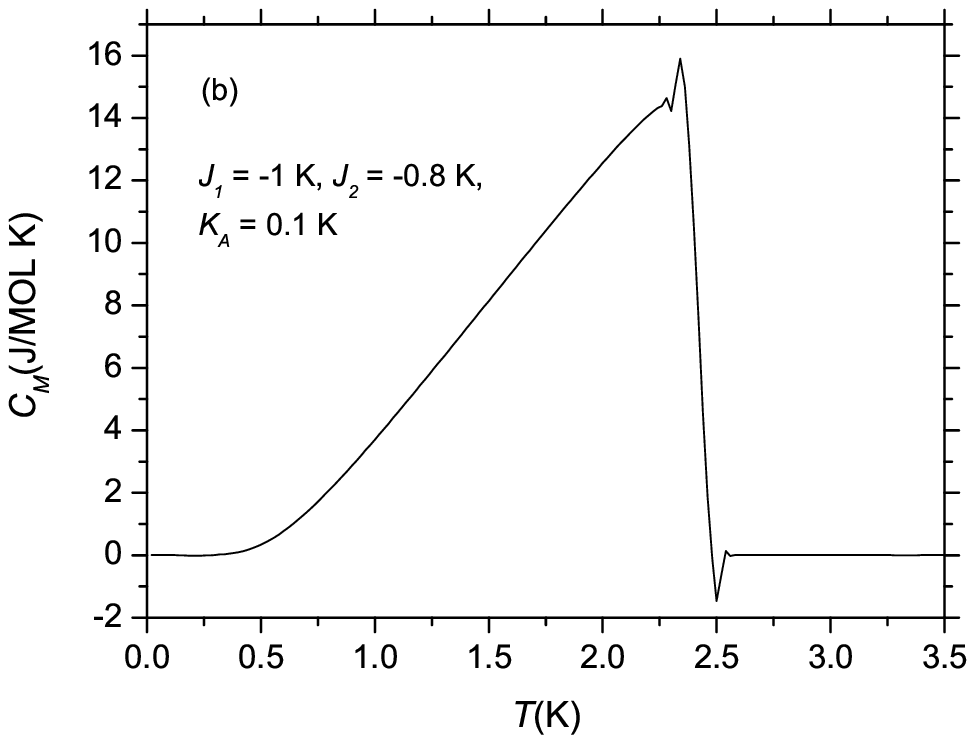,width=.45\textwidth,height=6.5cm,clip=}} 

\centering{}%
\parbox[c]{16cm}{%
{\small{}{}\textbf{\small{}Figure 3.}{\small{} (a) Total energy and
total free energy, and (b) specific heat, per mole of spins calculated
with the SCA approach for the antiferromagnetic nanotube in the absence
of external magnetic field. Here $N_{L}$ = 30.}}%
} 
\end{figure*}

The total free energy $F$, total energy $E$, magnetic entropy $S_{M}$
and specific heat $C_{M}$ of these canonical systems can be evaluated
with 
\begin{eqnarray}
F=-k_{B}T\log Z_{N},\space\space & E=-\frac{\partial}{\partial\beta}\log Z_{N}\;,\nonumber \\
S_{M}=\frac{E}{T}+k_{B}\log Z_{N}, & C_{M}=T\left(\frac{\partial S_{M}}{\partial T}\right)_{B}\;,
\end{eqnarray}
successively, where $\beta=1/(k_{B}T)$ and $Z_{N}$ is the partition
function of the whole system. Figure 3(a,b) display the $F$, $E$,
and $C_{M}$ curves obtained with the SCA approach for the antiferromagnetic
nanotube. It is very interesting to find that for the counterparting
ferromagnetic tube, i.e., as ${\cal J}_{1}$ = 1 K and ${\cal J}_{2}$
= 0.8 K, the calculated $F$, $E$ and $C_{M}$ curves are almost
exactly identical. The sudden change of $E$ and the sharp peak in
the $C_{M}$ curve near $T_{N}$ are the signs of phase transition.

%%%%%%%%%Fig4%%%%%%%%%%%%%%%
\begin{figure*}[ht]
\centerline{\epsfig{file=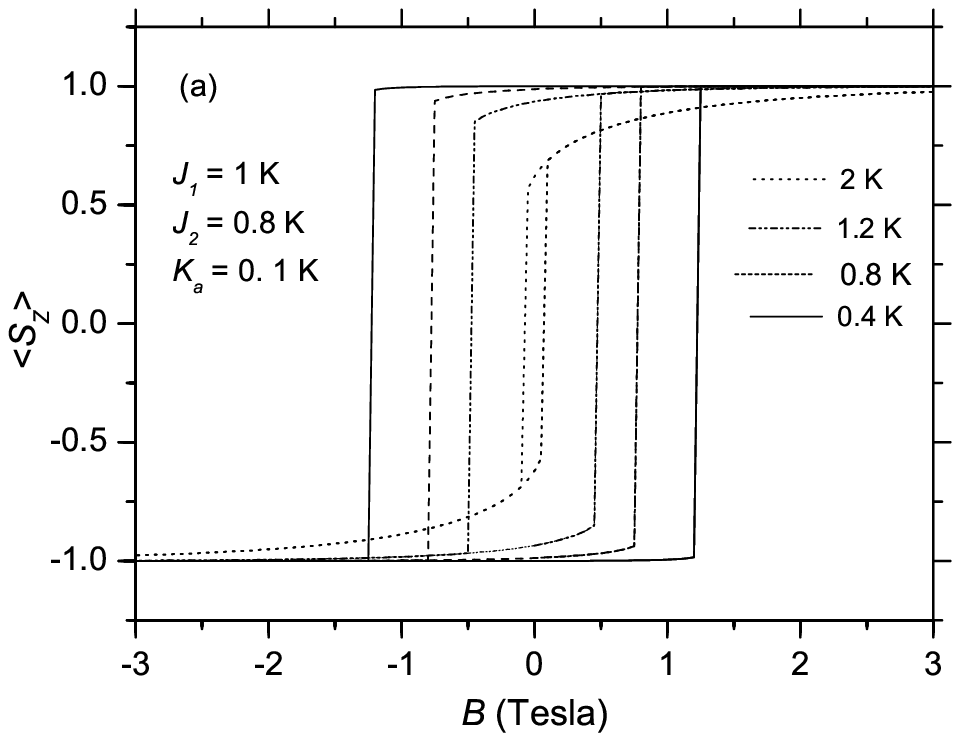,width=.45\textwidth,height=6.5cm,clip=}
\epsfig{file=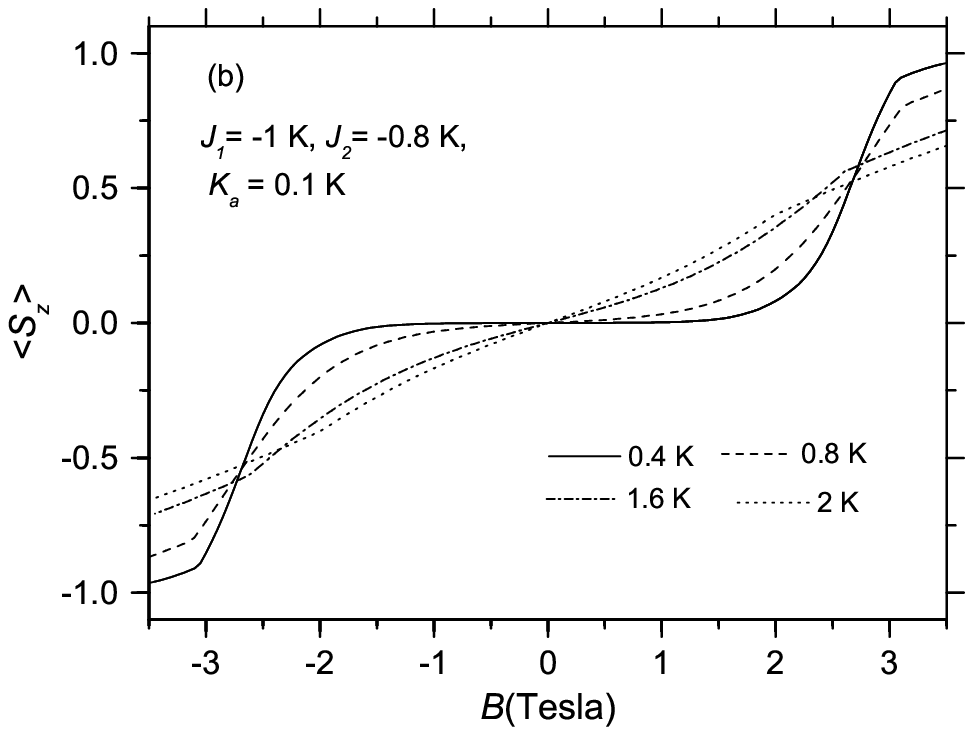,width=.45\textwidth,height=6.5cm,clip=}} 

\centering{}%
\parbox[c]{16cm}{%
{\small{}{}\textbf{\small{}Figure 4.}{\small{} Hysteresis curves
of the (a) ferromagnetic, and (b) antiferromagnetic nanotubes calculated
with the SCA approach at different temperatures. Here $N_{L}$ = 30.
}}%
} 
\end{figure*}

To learn how these two sorts of nanosystems react to the external
magnetic field, we then studied their hysteresis processes with the
SCA approach. Fig.4(a,b) show the longitudinal hysteresis curves calculated
at four different temperatures below $T_{M}$ with the SCA approach
for the ferromagnetic and antiferromagnetic nanotubes, respectively.
In the ferromagnetic case, hysteresis loops are formed at all four
temperatures. Especially at $T$ = 0.4 K, the loop is almost a rectangle.
If the external magnetic field is applied antiparallel to the original
magnetization, all spins are suddenly rotated for 180$^{0}$ at $B=B_{c}$=
1.25 T for instance. In the antiferromagnetic case, no hysteresis
loop can be formed as shown in Fig.4(b). And in the both cases, $\langle S_{x}\rangle$
and $\langle S_{y}\rangle$ always vanish if the magnetic field is
exerted exactly in the $z$-direction. In our previous SCA simulations
for an antiferromagnetic nanoparticle, it was discovered there that
a strong external magnetic field, applied in the direction along which
the antiferromagnetically coupled spins ordered inside the core region,
could rotate them off their original direction, forming symmetric
pattern around the field direction \cite{liuphye12}. The geometrical
shape of the nanotubes has indeed endowed them with very typical ferromagnetic
and antiferromagnetic characters.

\section{One dimensional ferromagnetic and antiferromagnetic chain models}

As just described, when a weak uniaxial anisotropy along the longitudinal
axis is present, all spins order spontaneously in the easy $z$-direction.
Even when an external magnetic field is applied along the central
longitudinal axis, nonzero $\langle S_{x}\rangle$ and $\langle S_{y}\rangle$
could not be observed. These peculiar features allow us to build up
one dimensional ferromagnetic and antiferromagnetic chain models.
In the ferromagnetic case, there are $N_{L}$ equally spaced spins
on the line along the longitudinal axis. The $m$-th spin $S_{z}(m)$
interacts with its left neighbor $S_{z}(m-1)$ and its right neighbor
$S_{z}(m+1)$ with strength ${\cal J}_{1}$, other two neighbors $S_{z}(m)$
on the same circle around the central axis with strength ${\cal J}_{2}$.
Since the spins only order in the $z$-direction, we then have, for
the Hamiltonian given in Eq.(\ref{hamlt}), three pure eigenvectors
$|\psi_{1}\rangle=|+1\rangle$, $|\psi_{2}\rangle=|0\rangle$, and
$|\psi_{3}\rangle=|-1\rangle$ for the $m$-th spin, and the corresponding
eigenvalues are expressed as 
\begin{eqnarray}
\varepsilon_{1}= & -\left[{\cal J}_{1}\left(\langle S_{z}(m-1)\rangle+\langle S_{z}(m+1)\rangle\right)+2{\cal J}_{2}\langle S_{z}(m)\rangle+g_{S}\mu_{B}B\right]-K_{A}\;,\nonumber \\
\varepsilon_{2}= & 0\;,\\
\varepsilon_{3}= & \left[{\cal J}_{1}\left(\langle S_{z}(m-1)\rangle+\langle S_{z}(m+1)\rangle\right)+2{\cal J}_{2}\langle S_{z}(m)\rangle+g_{S}\mu_{B}B\right]-K_{A}\;,\nonumber 
\end{eqnarray}
respectively. This chain has a finite length $N_{L}$. However, if
we let $0\leq m\leq N_{L}+1$, $\langle S_{z}(0)\rangle$ and $\langle S_{z}(N_{L}+1)\rangle$
assigned to 0 at the beginning, thus all spins can be treated equally.
Further assuming $\xi={\cal J}_{1}\left(\langle S_{z}(m-1)\rangle+\langle S_{z}(m+1)\rangle\right)+2{\cal J}_{2}\langle S_{z}(m)\rangle+g_{S}\mu_{B}B$,
we finally get a formula 
\begin{equation}
\langle S_{z}(m)\rangle=\frac{2\exp{(K_{A}/k_{B}T)}\sinh{(\xi/k_{B}T)}}{1+2\exp{(K_{A}/k_{B}T)}\cosh{(\xi/k_{B}T)}}\;,
\end{equation}
by making use of quantum theory.

An antiferromagnet can be considered to be composed of two oppositely
oriented $A$ and $B$ subsystems. Thus, in the simplified model,
an $A$ type spin interacts only with the nearest $B$-type spins,
and vice versa. More specifically, an $m$-th $A$ spin $S_{z}^{A}(m)$
interacts with a $B$-type neighbor $S_{z}^{B}(m-1)$ on its left,
another $B$-type neighbor $S_{z}^{B}(m+1)$ on its right, both with
strength ${\cal J}_{1}$, and other two $B$ type neighbors $S_{z}^{B}(m)$
belonging to the same circle around the central axis with strength
${\cal J}_{2}$, and vice versa. Thus, for the same sake just described,
we have once again three pure eigenvectors $|\psi_{1}^{A,B}\rangle=|+1\rangle$,
$|\psi_{2}^{A,B}\rangle=|0\rangle$, and $|\psi_{3}^{A,B}\rangle=|-1\rangle$
for the $m$-th $A$ or $B$ spin, respectively, and the corresponding
eigenvalues are given by 
\begin{eqnarray}
\varepsilon_{1}^{A,B}= & -\left[{\cal J}_{1}\left(\langle S_{z}^{B,A}(m-1)\rangle+\langle S_{z}^{B,A}(m+1)\rangle\right)+2{\cal J}_{2}\langle S_{z}^{B,A}(m)\rangle+g_{S}\mu_{B}B\right]-K_{A}\;,\nonumber \\
\varepsilon_{2}^{A,B}= & 0\;,\\
\varepsilon_{3}^{A,B}= & \left[{\cal J}_{1}\left(\langle S_{z}^{B,A}(m-1)\rangle+\langle S_{z}^{B,A}(m+1)\rangle\right)+2{\cal J}_{2}\langle S_{z}^{B,A}(m)\rangle+g_{S}\mu_{B}B\right]-K_{A}\;,\nonumber 
\end{eqnarray}
respectively. Assuming $\eta_{A,B}={\cal J}_{1}\left(\langle S_{z}^{A,B}(m-1)\rangle+\langle S_{z}^{A,B}(m+1)\rangle\right)+2{\cal J}_{2}\langle S_{z}^{A,B}(m)\rangle+g_{S}\mu_{B}B$,
and making use of quantum theory, one can easily deduce a pair of
coupled formulas 
\begin{eqnarray}
\langle S_{z}^{A}(m)\rangle=\frac{2\exp{(K_{A}/k_{B}T)}\sinh{(\eta_{B}/k_{B}T)}}{1+2\exp{(K_{A}/k_{B}T)}\cosh{(\eta_{B}/k_{B}T)}}\;\nonumber \\
\langle S_{z}^{B}(m)\rangle=\frac{2\exp{(K_{A}/k_{B}T)}\sinh{(\eta_{A}/k_{B}T)}}{1+2\exp{(K_{A}/k_{B}T)}\cosh{(\eta_{A}/k_{B}T)}}\;,
\end{eqnarray}
for the two sorts of oppositely oriented spins.

%%%%%%%%%Fig5%%%%%%%%%%%%%%%
\begin{figure*}[ht]
\centerline{\epsfig{file=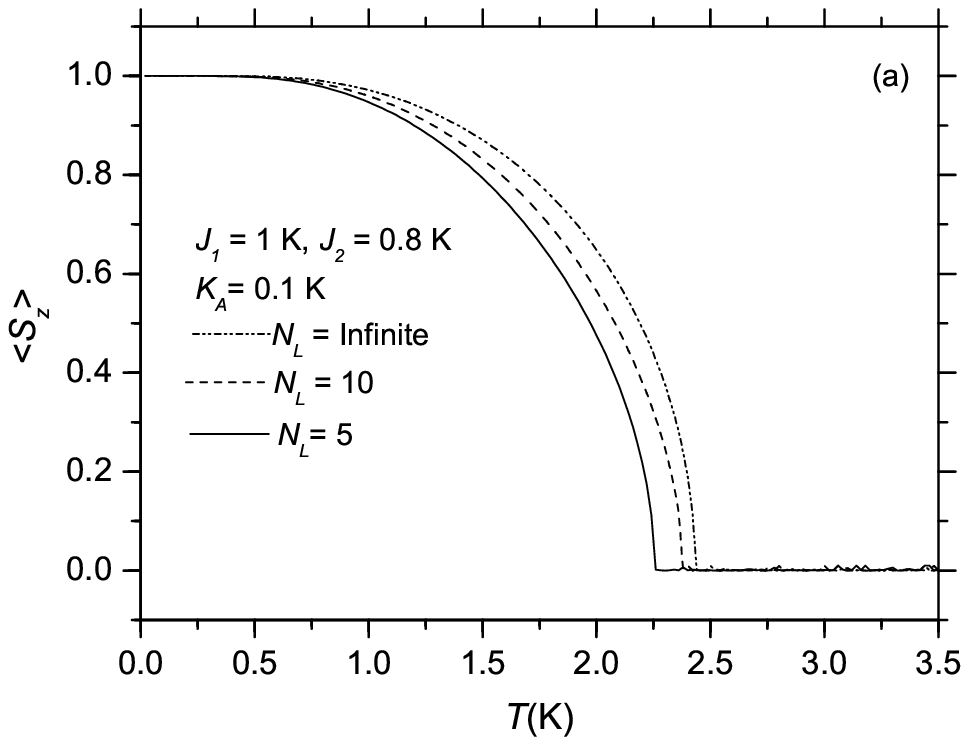,width=.45\textwidth,height=6.5cm,clip=}
\epsfig{file=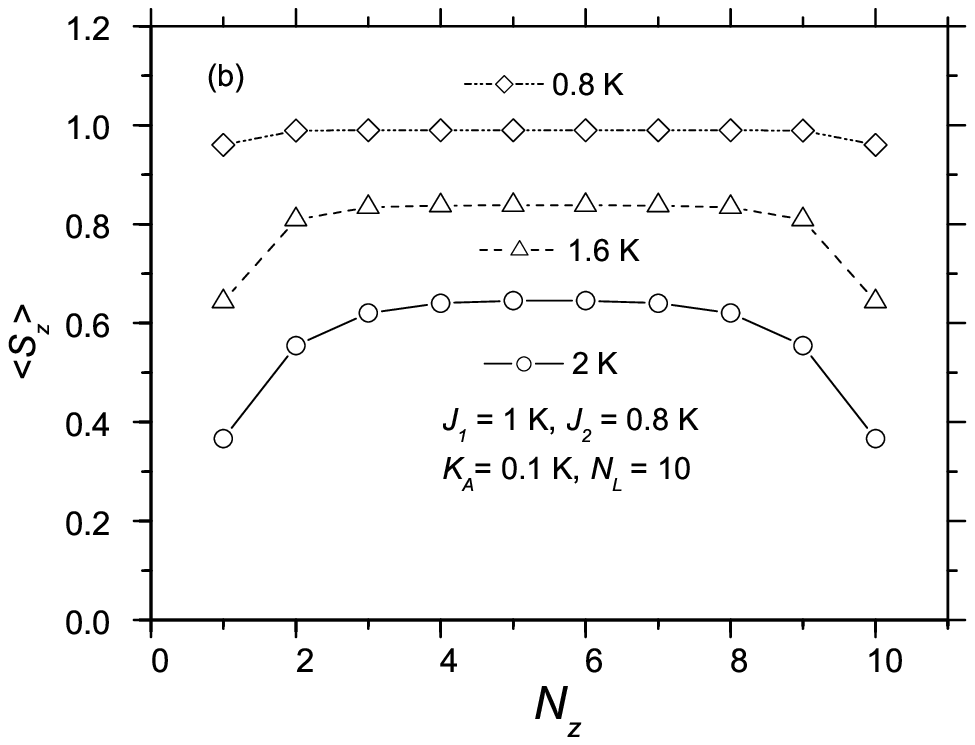,width=.45\textwidth,height=6.5cm,clip=}} 

\centering{}%
\parbox[c]{16cm}{%
{\small{}{}\textbf{\small{}Figure 5.}{\small{} Calculated $\langle S_{z}\rangle$
(a) as the functions of temperature for the ferromagnetic nanotubes
with different lengths, (b) as the functions of $N_{z}$ at different
temperatures for a ferromagnetic tube of length $N_{L}$ = 10, by
means of the one-dimensional ferromagnetic chain model in the absence
of external magnetic field. }}%
} 
\end{figure*}

\section{Results obtained with the one dimensional magnetic chain models}

Using the one-dimensional chain models, we have successfully reproduced
the results given in Fig.(1,2,4), no matter if the external magnetic
field was absent or applied in the $z$ direction. There is not need
to present those results here once again. Obviously, computational
speed can be greatly accelerated by using these theoretical models
because of their simplicity in comparison with the SCA approach.

With the two theoretical models, it is much easy to focus our studies
on the finite size effects. If the periodic condition is applied,
we then have an infinitely long nanotube. Fig.5(a) depicts the spontaneous
magnetization curves calculated with the one-dimensional chain model
for three ferromagnetic nanotubes of finite and infinite lengths,
respectively. We find here that a short tube has a low Curie temperature
$T_{C}$, however the length does not strongly affect the $T_{C}$
value. To fully understand the size effects, we plot the magnetization
curves for the tube with a length of $N_{L}$ = 10 at three different
temperatures in Fig.5(b). Near the two ends, the spins there have
weaker magnitudes than those in the middle region; and at higher temperatures,
these effects become much stronger. Fig.5(b) looks very similar to
Fig.1(d). But now, because of the short length, the size effects,
for example at $T$ = 2 K£\textlnot{} spread deeply into the very
middle part of the nanotube. %%%%%%%%%Fig6%%%%%%%%%%%%%%%
\begin{figure*}[ht]
\centerline{\epsfig{file=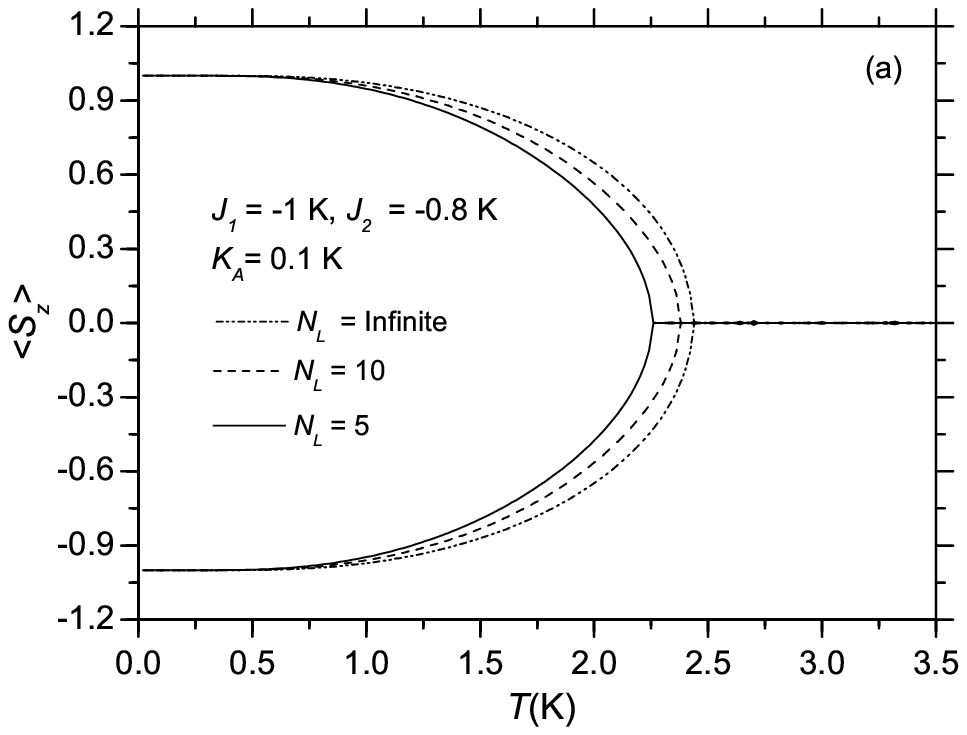,width=.45\textwidth,height=6.5cm,clip=}
\epsfig{file=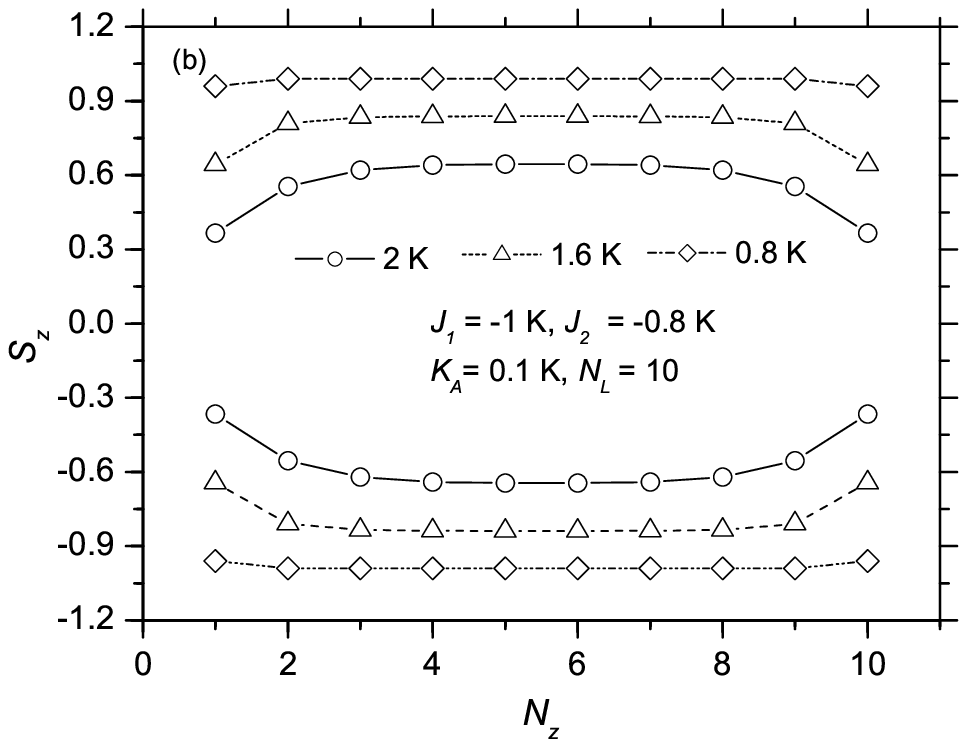,width=.45\textwidth,height=6.5cm,clip=}} 

\centering{}%
\parbox[c]{16cm}{%
{\small{}{}\textbf{\small{}Figure 6.}{\small{} Calculated $\langle S_{z}\rangle$
(a) as the functions of temperature for the antiferromagnetic nanotubes
with different lengths, (b) as the functions of $N_{z}$ at different
temperatures for an antiferromagnetic tube with $N_{L}$ = 10, by
means of the one-dimensional antiferromagnetic chain model in the
absence of external magnetic field. }}%
} 
\end{figure*}

%%%%%%%%%Fig7%%%%%%%%%%%%%%%
\begin{figure*}[ht]
\centerline{\epsfig{file=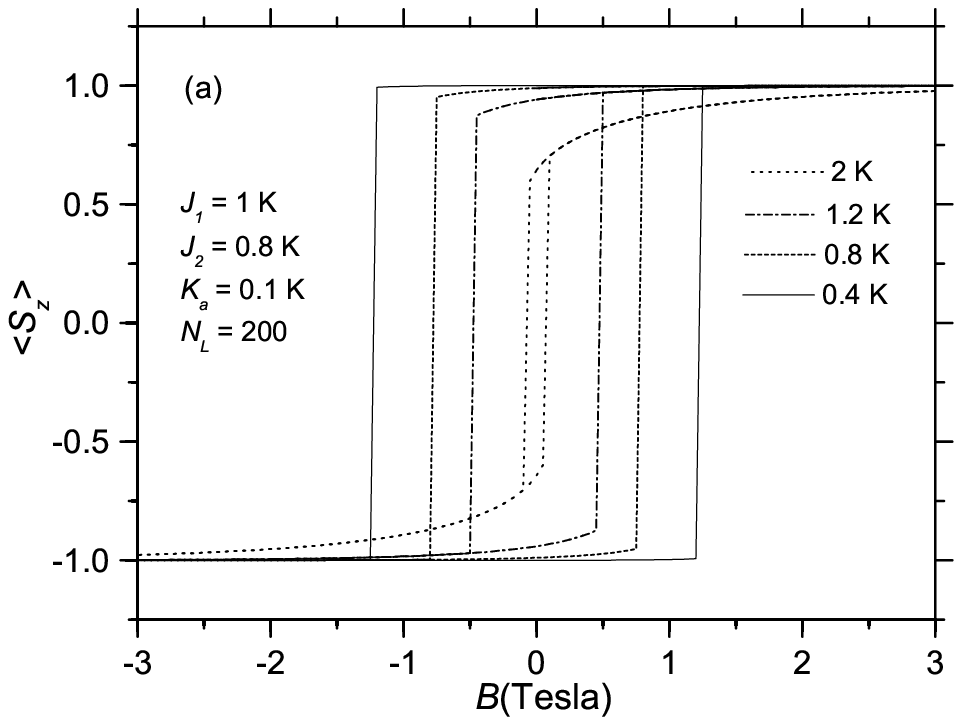,width=.45\textwidth,height=6.5cm,clip=}
\epsfig{file=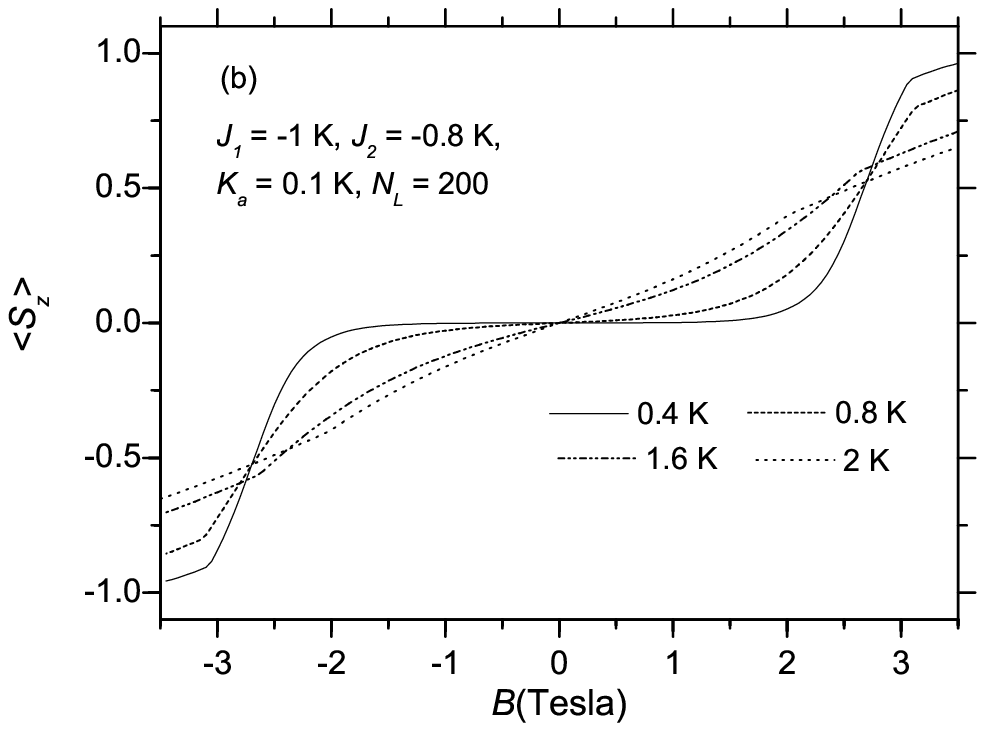,width=.45\textwidth,height=6.5cm,clip=}} 

\centering{}%
\parbox[c]{16cm}{%
{\small{}{}\textbf{\small{}Figure 7.}{\small{} Calculated hysteresis
curves of the (a) ferromagnetic, and (b) antiferromagnetic nanotubes
at different temperatures by means of the one-dimensional ferromagnetic
and antiferromagnetic chain models. Here $N_{L}$ = 200. }}%
} 
\end{figure*}

For comparison, the magnetization curves for the three antiferromagnetic
tubes, as displayed in Fig.6, were calculated with the one-dimensional
chain model by only changing the signs of ${\cal J}_{1}$ and ${\cal J}_{2}$
used in Fig(5) . Now we find the similar size effects as just described
for the ferromagnetic nanotubes. And more interestingly, if we only
draw the upper parts of Fig.6(a,b), the curves will coincide exactly
with their counterparts depicted in Fig.5(a,b) correspondingly. Therefore,
we can say the two sorts of systems are symmetric, and our calculated
results are self-consistent.

The relatively faster computational speed of the theoretical models
makes it easy for us to do simulations for long nanotubues. Fig.7(a,b)
show the longitudinal hysteresis curves of the ferromagnetic and antiferromagnetic
nanotubes of a length $N_{L}$ =200, obtained by means of the theoretical
models using the parameters given herein. The curves shown here for
the long nanotubes actually differ slightly from those shown in Fig.4(a,b),
even though their lengths differ four times. Once again, the nanosystems
exhibit typical ferromagnetic and antiferromagnetic characters.

%%%%%%%%%Fig8%%%%%%%%%%%%%%%
\begin{figure*}[ht]
\centerline{\epsfig{file=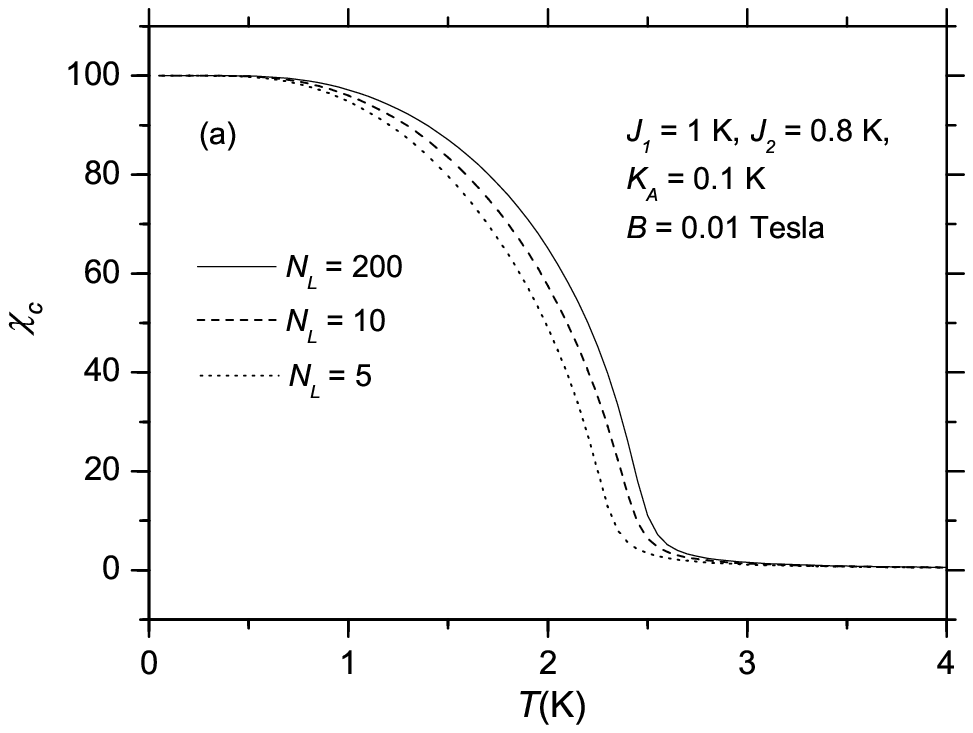,width=.45\textwidth,height=6.5cm,clip=}
\epsfig{file=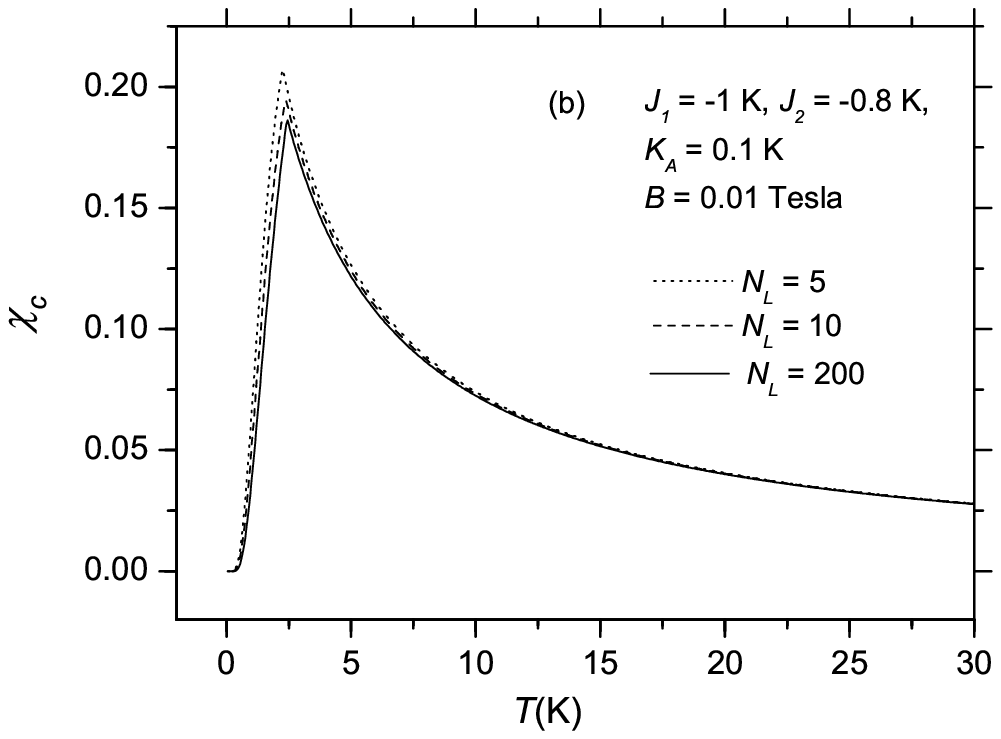,width=.45\textwidth,height=6.5cm,clip=}} 

\centering{}%
\parbox[c]{16cm}{%
{\small{}{}\textbf{\small{}Figure 8.}{\small{} Calculated longitudinal
susceptibility curves for (a) a ferromagnetic, and (b) an antiferromagnetic
nanotubes with different lengths, by means of the one-dimensional
ferromagnetic and antiferromagnetic chain models, respectively. }}%
} 
\end{figure*}

Finally, we calculated the susceptibilities of the two sorts of nanosystems.
For the purpose, a very weak external magnetic field is usually considered
to be exerted in a special direction, and the susceptibility can be
approximately estimated as the ratio of the induced/enhanced magnetization
to the applied magnetic field. So, we defined the longitudinal susceptibility
as $\chi_{c}\approx\langle S_{z}\rangle/B$, and a field of 0.01 Tesla
was assumed to be applied along the $z$-axis to calculate the quantities
with the one-dimensional magnetic chain models. The unit of the $\chi_{c}$
is 1/(Tesla Per spin). For simplicity, it is omitted in the figures.

Fig.8(a) displays the calculated results for three ferromagnetic nanotubes
of different lengths. At very low temperatures, $\langle S_{z}\rangle$
is close to the saturated value $S=1$, thus $\chi_{c}\approx\langle S_{z}\rangle/B\approx100$/(Tesla
Per spin). As temperature arises, $\chi_{c}$ attenuates. The applied
magnetic field modify each curve's shape around $T_{C}$ considerably
so that they attenuate continuously. Due to the size effects, we can
observe here three distinct curves.

In the antiferromagnetic case, as $T$ arises below $T_{N}$, the
applied magnetic field gradually rotate those magnetic moments antiparallel
to it to its direction. So in Fig.8(b), we observe that $\chi_{c}$
increases until $T_{N}$, where a peak appears. But above $T_{N}$,
the disordering effects of temperature becomes stronger and stronger
gradually, so we find all $\chi_{c}$'s fade with increasing temperature.
Again, we can see three curves near $T_{N}$'s, which is the evidence
of size effects.

\section{Conclusions and Discussion}

We have also performed simulations by exchanging the magnitudes of
${\cal J}_{1}$ and ${\cal J}_{2}$, that is, by assigning ${\cal J}_{1}=\pm$0.8
K and ${\cal J}_{2}=\pm$ 1 K, respectively, but we obtained the same
results as those shown in Fig.(1-4). This is easy to understand. For
instance, when ${\cal J}_{1}$ = 0.8 K and ${\cal J}_{2}$ = 1 K,
a spin still interacts with two spins with a strength of 1 K, other
two spins with a strength of 0.8 K, as in the case when ${\cal J}_{1}$
= 1 K and ${\cal J}_{2}$ = 0.8 K. So the calculated results with
the two sets of parameters are naturally same.

In summary, we firstly used the SCA approach to simulate the magnetic
structures for the two sorts of magnetic nanotubes, and calculated
their spontaneous magnetization, longitudinal hysteresis curves and
thermodynamic quantities. We found that no matter the external magnetic
field was absent, or applied along the central longitudinal axis,
the spins always aligned in that direction, seeming all bundled to
the tubes' surface because of their peculiar geometric structure,
so that the tubes exhibit typical ferromagnetic or antiferromagnetic
features, which may find applications in industry. In contrast, an
external magnetic field, applied in the direction along which the
antiferromagnetically coupled spins align inside a nanoparticle, is
able to turn the spins off the line to form symmetric pattern around
the field direction \cite{liuphye12}. Above findings in numerical
simulations enable us to build up one-dimensional ferromagnetic and
antiferromagnetic chain models, which could reproduce all the results
obtained in SCA simulations in principle, and were then employed to
investigate the size effects, and investigate the magnetic properties
of long tubes. Because of the simplicity of the theoretical models,
computational speed can be considerably improved. Especially, the
exact agreements between our numerical calculations and theoretical
analysis justify the correctness of our quantum simulation approach
once again.

\vspace{0.5cm}
 \centerline{\textbf{Acknowledgements} }Z.-S. Liu acknowledges the
financial support by National Natural Science Foundation of China
under grant No.~11274177. H. Ian is supported by the FDCT of Macau
under grant 013/2013/A1, University of Macau under grants MRG022/IH/2013/FST
and MYRG2014-00052-FST, and National Natural Science Foundation of
China under Grant No.~11404415.

\end{document}